\documentclass[twocolumn,showpacs,groupedaddress]{revtex4}
\usepackage{graphicx}
\usepackage{epsfig}
\usepackage{subfigure}
\usepackage{amsmath}

\begin{document}

\newcommand{\ket}[1]{|#1\rangle}
\newcommand{\bra}[1]{\langle#1|}
\newcommand{\Tr}{\text{Tr}}

\title{
Lyapunov Control on  Quantum Open System in Decoherence-free
Subspaces}
\author{ W. Wang, L. C. Wang, X. X. Yi}

\affiliation{School of Physics and Optoelectronic Technology, Dalian
University of Technology, Dalian 116024, China}

\begin{abstract}
A scheme  to drive and manipulate a finite-dimensional quantum
system in the decoherence-free subspaces(DFS) by Lyapunov control is
proposed. Control fields are established by Lyapunov function. This
proposal can drive the open quantum system into the DFS and
manipulate it to any desired eigenstate of the free Hamiltonian. An
example which consists of a four-level system with three long-lived
states driven by two lasers is presented to exemplify  the scheme.
We have performed numerical simulations for the dynamics of the
four-level system, which show that the scheme works good.
\end{abstract}

 \maketitle

\section{introduction}

Manipulating  the time evolution of a quantum system is a major task
required for quantum information processing. Several strategies for
the control of a quantum system have been proposed in the past
decade\cite{dong09}, which can be divided into coherent and
incoherent control, according to how the controls enter the
dynamics.  Among the quantum control strategies, Lyapunov control
plays an important role in quantum control theory. Several papers
have be published recently to discuss the application of Lyapunov
control to quantum
systems\cite{vettori02,mirrahimi05,altafini07,yi09}. Although the
basic mathematical formulism for Lyapunov control is well
established, many questions remain when one considers its
applications in quantum information processing, for instance,  the
Lyapunov  control on open system and the state manipulation in its
decoherence-free subspace.

As a collection of states that  undergo unitary evolution in the
presence of decoherence, the decoherence-free subspaces
(DFS)~\cite{zanardi97} and noiseless subsystem(NS)\cite{knill00} are
promising concept in quantum information processing. Experimental
realizations of DFS have been achieved with photons~\cite{kwiat00}
and in nuclear spin systems~\cite{viola01}. A decoherence-free
quantum memory for one qubit has been realized experimentally with
two trapped ions~\cite{kielpinski01,langer05}. An in-depth study of
quantum stabilization problems for DFS and NS of Markovian quantum
dynamics was presented in\cite{ticozzi08}.

Most recently, we have proposed a scheme to drive an open quantum
system into the decoherence-free subspaces\cite{yi09}. This scheme
works also for closed quantum system, by replacing the DFS with a
desired subspace. The result suggests that it is possible to drive a
quantum system to a set of states (for example, the DFS in the
paper), however it is difficult to manipulate the system into a
definite quantum state in the DFS. The aim of this paper is to
design a Lyapunov control to drive an open system to a definite
state in the DFS. The Lyapunov control has been proven to be a
sufficient  simple control to be analyzed rigorously, in particular,
the control can be shown to be highly effective for systems that
satisfy certain sufficient conditions, which roughly speaking are
equivalent to the controllability of the linearized system. In
Lyapunov control, Lyapunov functions which were originally used in
feedback control to analyze the stability of the control system,
have formed the basis for new control design. By properly choosing
the Lyapunov function, our analysis and numerical simulations show
that the control scheme  works good.

This paper is organized as follows. In Sec. \ref{gf}, we present a
general analysis of Lyapunov control for open quantum systems,
Lyapunov functions and control fields  are given and discussed. To
illustrate the general formulism, we exemplify a four-level system
with 2-dimensional DFS in Sec. \ref{exa}, showing that  the system
can be controlled to a desired state  in the DFS  by  Lyapunov
control. Finally, we conclude our results in Sec. \ref{con}.

\section{general formulism}\label{gf}
We can model a controlled quantum system  either by a closed system,
or by an open system governed by a master equation. In this paper,
we restrict our discussion to a $N$-dimensional open quantum system,
and consider its dynamics  as Markovian and therefore the dynamics
obeys the Markovian master equation ($\hbar=1,$ throughout this
paper),
\begin{eqnarray}
\dot{\rho}=-i[H,\rho]+{\mathcal L}(\rho), \label{mef}
\end{eqnarray}
where ${\mathcal L}(\rho)=\frac 1 2\sum_{m=1}^M \lambda_m([L_m,\rho
L_m^{\dagger}]+[L_m\rho,L_m^{\dagger}]),$ $ H=H_0+\sum_{n=1}^F
f_n(t)H_n$.  $\lambda_m (m=1,2,...,M)$ are positive and
time-independent parameters, which characterize the decoherence.
$L_m (m=1,2,...,M)$ are jump operators. $H_0$ is a free Hamiltonian
and $H_n (n=1,2,...,F)$ are control Hamiltonian, while $f_n(t)
(n=1,2,...,F)$ are control fields. Equation (\ref{mef}) is of
Lindblad form, this means that the solution to Eq. (\ref{mef}) has
all the required properties of a physical density matrix at all
times.

By  definition, DFS is composed  of states that undergo unitary
evolution. Considering the fact that there are many ways for a
quantum system to evolve unitarily, we  focus on the DFS here that
the dissipative part ${\mathcal {L}(\rho)}$ of the master equation
is zero, leading to the following conditions for
DFS\cite{karasik08}. A space spanned by $\mathcal{H}_{DFS}=\{
|\psi_1\rangle, |\psi_2\rangle, ..., |\psi_D\rangle \}$ is a
decoherence-free subspace for all time $t$ if and only if (1)
$\mathcal{H}_{DFS}$ is invariant under $H_0$; (2)
$L_m|\psi_n\rangle=c_m|\psi_n\rangle$ and (3)
$\Gamma|\psi_n\rangle=g|\psi_n\rangle$ for all $n=1,2,...,D$ and
$m=1,2,...,M$ with $g=\sum_{l=1}^M\lambda_l|c_l|^2,$ and
$\Gamma=\sum_{m=1}^M\lambda_mL_m^{\dagger}L_m.$ With these
notations, the goal of this paper can be formulated as follows. We
wish to apply a specified set of control fields $\{ f_i(t),
n=1,2,...,F\}$ in Eq. (\ref{mef}) such that $\rho(t)$ evolves into a
desired state in the DFS and stays there forever. In contrast to the
conventional control problem\cite{wangx09}, we here develop the
control strategy to open system.

We use
\begin{equation}
V(\rho)=\Tr(\rho\hat{A})
\end{equation}
as a Lyapunov function, where $\hat{A}$ is hermitian and
time-independent. First, we analyze the structure of critical points
for $V(\rho)$ with restriction $\Tr(\rho)=1.$ To determine the
structure of $V(\rho)$ around one of its critical points, for
example $\rho_c=\sum_j p_j^c|A_j\rangle\langle A_j|$ , we consider a
finite variation $\delta \rho$ such that $\Tr(\rho_c+\delta
\rho)=1.$ Here we denote the normalized eigenvectors and eigenvalues
of $\hat{A}$ by $|A_i\rangle$ and $A_i$ ($i=1,2,3,...,N$),
respectively. Express $(\rho_c+\delta \rho)$ in the basis of the
eigenvectors of $\hat{A},$
\begin{eqnarray}
&&\rho_c+\delta\rho=\sum_j p_j^c|A_j+\delta A_j\rangle\langle
A_j+\delta A_j|,\nonumber\\
&&|A_j+\delta A_j\rangle =
|A_j\rangle+\sum_{\alpha=1}^N\delta_{\alpha}^j|A_{\alpha}\rangle.\label{expan}
\end{eqnarray}
The normalization condition $\Tr(\rho_c+\delta \rho)=1$ follows,
$$ \sum_j p_j^c(\delta_j^{j*}+\delta_j^j)+\sum_j
p_j^c\sum_{\alpha}\delta^{j*}_{\alpha}\delta^{j}_{\alpha}=0.$$ Then
\begin{equation}
V(\rho_c+\delta \rho)-V(\rho_c)=\sum_jp_j^c\sum_{\alpha\neq
j}(A_{\alpha}-A_j)\delta_{\alpha}^{j*}\delta_{\alpha}^j.
\end{equation}
Considering $\delta_{\alpha}^j$ as variation parameters and noting
$\delta_{\alpha}^{j*}\delta_{\alpha}^j\geq 0,$ we find that the
structure of $V(\rho)$ around the critical point $\rho_c$ depends on
the ordering of the eigenvalues: $\rho_c$ is a local maximum as a
function of the variations $\delta_{\alpha}^j$ if and only if $A_j$
is the largest eigenvalue, a local minimum iff $A_j$ is the smallest
eigenvalue and a saddle point otherwise. This observation leads us
to suspect that the minimum of $V$ is asymptotically attractive, in
other words, the control field based on this Lyapunov function would
drive the open system to the eigenstate of $\hat{A}$ with the
smallest eigenvalue. We will show through an example that this is
exactly the case.

Now we establish the control fields $f_n(t).$ $V(\rho)=\Tr(\rho
\hat{A})$ yields,
$$\dot{V}=\Tr({\mathcal
L}(\rho) \hat{A})-i\Tr(\rho[\hat{A},\sum_n f_n(t) H_n]),$$ where we
choose $[\hat{A},H_0]=0,$ because ($\hat{a}$, $\hat{b}$ any
operators) $\Tr[\hat{a},\hat{b}]=0,$ i.e., the commutator can never
be sign definite. The choice of $[\hat{A},H_0]=0$ implies that $H_0$
and $\hat{A}$ must have the same eigenvectors, then the control
field would drive the open system into an eigenstate of the
Hamiltonian $H_0$. To make $\dot{V}\leq 0,$ we choose a $f_{j_0}(t)$
such that
\begin{eqnarray}
&&f_{j_0}(t)=-i\frac{\Tr({\mathcal
L}(\rho)\hat{A})}{\Tr([\hat{A},H_{j_0}]\rho)},\nonumber\\
&&f_j(t)=-i\kappa_j(\Tr([\hat{A},H_j]\rho))^*, \ \ \mbox{for}\ \
j\neq j_0.\label{fs}
\end{eqnarray}
Here $\kappa_j>0$ will be refereed as the strength of the control.
Then the evolution of the open system with Lyapunov control can be
described by the following nonlinear equations
\begin{eqnarray}
\dot{\rho}(t)&=&-i[H_0+\sum_{n} f_n(t)H_n,
\rho(t)]+\mathcal{L}(\rho), \label{nle1}
\end{eqnarray}
where $f_n(t)$ is determined by Eq.(\ref{fs}). It should be
emphasized that $f_{j_0}$ always exists. To find $f_{j_0}$,
$\Tr([\hat{A},H_{j_0}]\rho)\neq 0$ is required. This can be done by
construction. Now we show that $f_{j_0}$ is real. By the definition
of $\mathcal {L} (\rho)$, $\mathcal {L} (\rho)$ is hermite, then
$\Tr({\mathcal L}(\rho)\hat{A})$ can be treated as the time
derivative of $\langle\hat{A}\rangle$ and thus is real. Identifying
$\hat{A}$ with a hermitian operator for a system described by the
Hamiltonian $H_{j_0}$, we have $i\frac{\partial \hat{A}}{\partial
t}=[\hat{A}, H_{j_0}],$ so $\Tr(i[\hat{A},H_{j_0}]\rho)$ is real. By
the same virtue, we can show that all the control fields are real as
long as the control Hamiltonian $H_j$ ($j=1,2,3...$) are hermitian.

 By the LaSalle's invariant
principle\cite{lasalle61}, the autonomous dynamical system
Eq.(\ref{nle1}) converges to an invariant set defined by
$\mathcal{E}=\{\dot{V}=0\}$. This set is in general not empty and of
finite dimension, indicating that it is easy to manipulate an open
system to a set of states but difficult to control it from an
arbitrary initial state to a given target state. Fortunately, by
elaborately designing the control Hamiltonian and the operator
$\hat{A}$, we can solve this problem as follows. The invariant set
defined by $\mathcal{E}=\{\dot{V}=0\}$ is an intersection of all
sets $\mathcal{E}_j$ $(j=1,2,3,...)$, each one satisfies,
$$\Tr(\hat{A}H_j\rho-H_j\hat{A}\rho)=0,$$
leading to $[\hat{A},\rho]=0$, $[\hat{A},H_j]=0$ or $[H_j,\rho]=0$.
By elaborately choosing $H_j$ ($j=1,2,3,...)$, we can set the
contribution of $[\hat{A},H_j]=0$ and $[H_j,\rho]=0$ to the
intersection (i.e., the invariant set $\mathcal{E}$) to zero. In
this case, the invariant set is a collection of state
$\{\rho_{in}\}$ that satisfies $[\hat{A},\rho_{in}]=0.$ Considering
that only the states in DFS are stable, we claim that we can
manipulate the system from any initial state to the target state in
DFS. In other words, we can design $\hat{A}$ such that
$\mathcal{E}\bigcap \mbox{DFS}$ contains only the target state.
 We emphasis that
although the control field $f_{j_0}(t)$ was specified to cancel
$\Tr(\mathcal{L}(\rho)\hat{A})$ in $\dot{V}$, it makes contribution
to the dynamics of the open system.

\section{example}\label{exa}
\begin{figure}
\includegraphics*[width=0.7\columnwidth,
height=0.5\columnwidth]{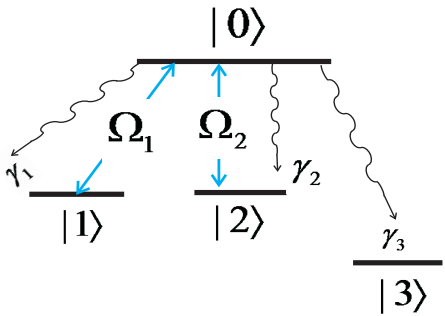}
 \caption{The schematic energy diagram. A four-level system with two degenerate stable states
 $|1\rangle$ and $|2\rangle$ in external laser fields.
 The two degenerate states are coupled to the excited state $|0\rangle$
 by two separate lasers with coupling constants $\Omega_1$ and $\Omega_2$, respectively. While the
 stable state $|3\rangle$ is isolated from the other levels. The excited state $|0\rangle$ decays to  $|j\rangle$
 ($j=1,2,3$) with decay rate $\gamma_j$.}\label{f1}
\end{figure}

As an example of the Lyapunov control strategy, we discuss below a
four-level system coupling to two external lasers, as shown in Fig.
\ref{f1},
The Hamiltonian of such a system has the form
\begin{equation}
H_0=\sum_{j=0}^2\Delta_j |j\rangle\langle j|+(\sum_{j=1}^2
\Omega_j|0\rangle\langle j|+h.c.),
\end{equation}
where $\Omega_j \  \ (j=1,2)$ are coupling constants. Without loss
of generality, in the following the coupling constants are
parameterized as $\Omega_1=\Omega \cos \phi$ and
$\Omega_2=\Omega\sin\phi$  with
$\Omega=\sqrt{\Omega_1^2+\Omega_2^2}.$  The excited state
$|0\rangle$ is not stable, it decays to the three stable states with
rates $\gamma_1$, $\gamma_2$ and $\gamma_3$, respectively. We assume
this process is Markovian and can be described by the Liouvillian,
\begin{equation}
\mathcal L(\rho)=\sum_{j=1}^3\gamma_j(\sigma_j^-\rho\sigma_j^+-\frac
1 2 \sigma_j^+\sigma_j^-\rho-\frac 1 2 \rho\sigma_j^+\sigma_j^-)
\end{equation}
with $\sigma_j^-=|0\rangle\langle j|$ and
$\sigma_j^+=(\sigma_j^-)^{\dagger}.$ It is not difficult to find
that the two degenerate dark states
\begin{eqnarray}
|D_1\rangle&=&\cos\phi|2\rangle-\sin\phi|1\rangle,\nonumber\\
|D_2\rangle&=&|3\rangle,
\end{eqnarray}
of the free Hamiltonian $H_0$ form a DFS. Now we show how to control
the system to a desired target state in  the DFS. For this purpose,
we choose the control Hamiltonian $H_c=\sum_{j=1}^3 f_j(t)H_{j}$
with
\begin{eqnarray}
&&H_1=\left(
\begin{array}{cccc}
1&1&1&1\\
1&1&1&1\\
1&1&1&1\\
1&1&1&1\\
\end{array}
\right),\nonumber\\
&& H_2=|D_1\rangle\langle D_2|+|D_2\rangle\langle
D_1|,\nonumber\\
&&H_3=|0\rangle\langle D_2|+|D_2\rangle\langle 0|.
\end{eqnarray}
We shall use Eq. (\ref{fs}) to determine the control fields
$\{f_n(t)\}$, and choose
\begin{eqnarray}
|\Psi\rangle&=&\sin\beta_1\cos\beta_3|0\rangle+\cos\beta_1\cos\beta_2|1\rangle\nonumber\\
&+& \cos\beta_1\sin\beta_2|2\rangle+\sin\beta_1\sin\beta_3|3\rangle
\label{inis}
\end{eqnarray}
as  initial states for the numerical simulation, where $\beta_1$,
$\beta_2$ and $\beta_3$ are allowed to change independently. We
should emphasis that it is difficult to exhaust all possible initial
states in the simulation, because for a 4-dimensional system, there
are 15 independent real parameters needed to describe a general
state, even for pure states, 6 real independent parameters are
required. The initial state written in Eq.(\ref{inis}) omits all
(three) relative phases between the  states $|0\rangle, |1\rangle,
|2\rangle$ and $|3\rangle$ in the superposition,  and satisfies the
normalization condition. $f_1(t)$ here is specified to cancel the
contribution of $\Tr [{\mathcal L}(\rho)\hat{A}]$ to $\dot{V},$ this
means that $f_1(t)=-i\frac{\Tr({\mathcal
L}(\rho)\hat{A})}{\Tr([\hat{A},H_{1}]\rho)},$ $f_2(t)$ and $f_3(t)$
are determined by Eq.(\ref{fs}).
\begin{figure}
\includegraphics*[width=0.7\columnwidth,
height=0.5\columnwidth]{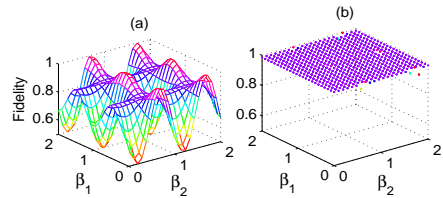} \vspace{-2cm}
\caption{Fidelity of the system in the target state $|D_1\rangle$
(a), and in the DFS
 (b). The control field $f_3(t)$ is turned off, i.e., $f_3(t)=0.$ $\Omega=5, \phi=\frac{\pi}{5},
 \beta_3=\frac{\pi}{3}, \gamma_1=\gamma_2=\gamma_3=\frac 1 3 \gamma, \kappa_2=1,$ $\Delta_0=4, \Delta_1=\Delta_2=2$
 and $
 \gamma=1.$}\label{f2}
\end{figure}

\begin{figure}
\includegraphics*[width=0.7\columnwidth,
height=0.5\columnwidth]{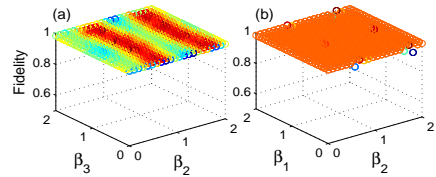}
 \vspace{-2cm}\caption{Fidelity of the control with $\hat{A}=|D_2\rangle\langle D_2|-|D_1\rangle\langle D_1|$
 (i.e., the target state is $|D_1\rangle$) as a function of the initial states. The parameters chosen are the
 same as Fig. \ref{f2}. The control strength $\kappa_2=1, \kappa_3=15$ were specified for this plot.}\label{f3}
\end{figure}

\begin{figure}
\includegraphics*[width=0.7\columnwidth,
height=0.5\columnwidth]{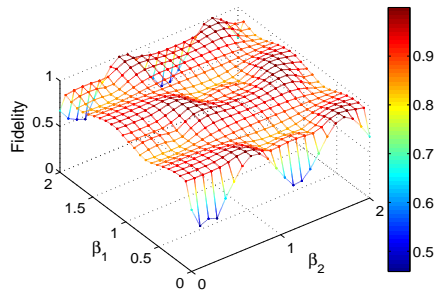}
 \caption{Fidelity of the control versus initial states. The target state is $|D_2\rangle$
 (or $\hat{A}=-|D_2\rangle\langle D_2|+|D_1\rangle\langle D_1|$). $\kappa_3=15$,
 the other parameters chosen are the same as in Fig.\ref{f2} are chosen
 for this plot. }\label{f4}
\end{figure}

We have performed extensive numerical simulation with the initial
states  Eq.(\ref{inis}). Numerical results are presented in Figs.
\ref{f2}-\ref{f4}. The control field $f_3(t)$ plays an important
role in this scheme as Fig. \ref{f2} shows. Fig. \ref{f2} tells us
that without the control field $f_3(t)$, the open system can be
driven into the DFS (with $\hat{A}$ given below), but it can not be
manipulated into a definite state in DFS.
The physics behind is the following. With the given $\hat{A}$ (see
below), $f_1(t)$ is always zero, so $H_1$ plays no role in the
control. The only control that enters the system is $f_2(t)H_2$.
From Eq.(\ref{fs}), we find that $f_2(t)$ takes zero provided
$\rho=x|D_2\rangle\langle D_2|+(1-x)|D_1\rangle \langle D_1|,$
 (where $x\geq 0)$, leading to the above observation.  When the control
field $f_3(t)$ is turned on. The four-level system can be controlled
to a desired state in DFS by properly choosing $\hat{A}.$
For example, when $\hat{A}=\hat{A}_1=|D_2\rangle\langle
D_2|-|D_1\rangle\langle D_1|,$ the system can be controlled into
$|D_1\rangle$ (see Fig.\ref{f3}), whereas $\hat{A}=-\hat{A}_1$ can
drive the system into $|D_2\rangle$ (see Fig.\ref{f4}). Based on the
formalism in Sec. \ref{gf}, $\hat{A}=\hat{A}_1$ together with the
controls could drive the system to the eigenstate of $\hat{A}_1$
with smallest eigenvalue (namely, $|D_1\rangle$), and to
$|D_2\rangle$ with $\hat{A}=-\hat{A}_1.$  As figure 4 shows,
however, the fidelity is not 1 for some initial states, for example
$\beta_1=0.$  The reason is as follows. Though  the choice of
$\hat{A}=-\hat{A}_1$ benefits the target state $|D_2\rangle$, since
$|D_2\rangle$ is the eigenstate of $\hat{A}$ with smallest
eigenvalue, the control $H_3=(|0\rangle\langle D_2|+h.c.)$ does not
favor the control target $|D_2\rangle$, because $H_3$ couples the
states $|0\rangle$ and $|D_2\rangle$, and $|0\rangle$ decays to the
three stable states equally. This observation suggests that
$h_3=(|0\rangle \langle D_1|+h.c.)$ instead of $H_3$ could help the
control when the target is $|D_2\rangle.$ Indeed further numerical
simulations confirm this prediction that the control $h_3$  can
drive the system into $|D_2\rangle$ with almost perfect fidelity
$99\%$.
\begin{figure}
\includegraphics*[width=0.7\columnwidth,
height=0.5\columnwidth]{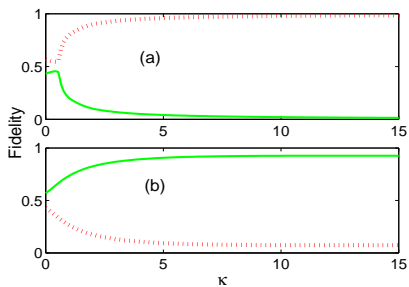}
 \caption{Fidelity as a function of the control strength $\kappa_3=\kappa$.
 (a)$\hat{A}=|D_2\rangle\langle D_2|-|D_1\rangle\langle D_1|,$
 (b)$\hat{A}=-|D_2\rangle\langle D_2|+|D_1\rangle\langle D_1|.$ $\phi=\frac\pi 4, \beta_1=\frac \pi 6,
 \beta_2=\frac \pi 3, \beta_3=\frac \pi 5,$ and $\kappa_2=1.$ The other parameters chosen are the same as Fig. \ref{f2}. }\label{f5}
\end{figure}
 The fidelity of the open system in the target state
depends on the strength $\kappa_3=\kappa$  of the control $f_3(t),$
the dependence is plotted in Fig.\ref{f5}. With large $\kappa$, the
system would asymptotically converge to the target state as
Fig.\ref{f5} shows. As expected, the control fields $f_2(t)$ and
$f_3(t)$ tend to zero when the open system converges to the target
state, see Fig.\ref{f6}
\begin{figure}
\includegraphics*[width=0.7\columnwidth,
height=0.5\columnwidth]{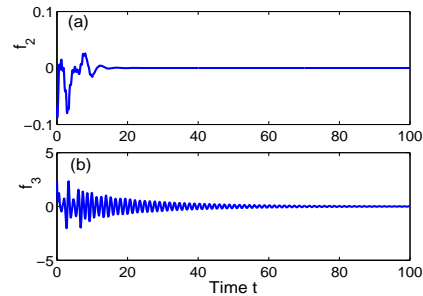}
 \caption{Control field $f_2(t)$ and $f_3(t)$ as a function of time.
 $\Omega=5, \phi=\frac\pi 5, \beta_1=\frac \pi 5,
 \beta_2=\frac \pi 4, \beta_3=\frac \pi 6, \kappa_2=1,$ and $\kappa_3=15.$
 $\hat{A}=|D_2\rangle\langle D_2|-|D_1\rangle\langle D_1|.$ $f_1(t)$ is zero in this scheme.}\label{f6}
\end{figure}

\section{conclusion}\label{con}
In summary, we have proposed a scheme to manipulate an open quantum
system in the decoherence-free subspaces. This study was motivated
by the fact that for Lyapunov control, it is usually difficult to
optimally control the system from an arbitrary initial state to a
given target state, this is due to the LaSalle's invariant
principle.  Our present study suggests that it is possible to drive
a quantum system to a desired state in DFS by elaborately designing
the controls. The results do not break the LaSalle's role, instead
it reduces the invariant set $\mathcal{E}$ to include the target
state only. To demonstrate the proposal we exemplify a four-level
system and numerically simulate the controlled dynamics. The
dependence of the fidelity on initial states as well as the control
fields are calculated and discussed. This scheme put the Lyapunov
control  on quantum open system one step forward, and shed light on
the quantum control in DFS.

\ \ \ \\
This work is supported by  NSF of China under grant Nos. 10775023
and 10935010.

\end{document}